\documentclass[lettersize,journal]{IEEEtran}
\usepackage{amsmath,amsfonts}
\usepackage{algorithmic}
\usepackage{algorithm}
\usepackage{array}
\usepackage[caption=false,font=footnotesize]{subfig}
\usepackage{textcomp}
\usepackage{url}
\usepackage{verbatim}
\usepackage{graphicx}
\usepackage{adjustbox}
\usepackage{hyperref}
\usepackage[noadjust]{cite}
\usepackage[svgnames,x11names,table,dvipsnames]{xcolor}
\usepackage[utf8]{inputenc}
\usepackage{pgfplots}
\usepgfplotslibrary{groupplots}
\pgfplotsset{compat=1.18}
\usepackage{tikz-3dplot}
\usepackage{pifont}
\usepackage{tikz}
\usetikzlibrary{arrows.meta, positioning}
\usetikzlibrary{shadings}
\usepackage{etoolbox}
\renewcommand{\baselinestretch}{0.99}
\makeatletter
\def\IEEEbibitemsep{0pt}
\makeatother
\usepackage{lipsum}

\usepackage{fancyhdr}

\begin{document}
\title{Parameterizing Operating-Point-Dependent IBR 
Using Coherent Operating Regions for
Sub-synchronous Oscillation Analysis}
\author{Gabriel Covarrubias Maureira,~\IEEEmembership{Member,~IEEE,} Balarko Chaudhuri,~\IEEEmembership{Fellow,~IEEE}, and Mark O'Malley,~\IEEEmembership{Fellow,~IEEE}
        % <-this % stops a space
\thanks{The authors are with the Department of Electrical and Electronic Engineering, Imperial College London.}
\thanks{This work was supported by the Engineering and Physical Sciences
Research Council grant number EP/Y025946/1 and by the Leverhulme International Professorship grant reference LIP-2020-002.}}
%\thanks{Manuscript received April 19, 2021; revised August 16, 2021.}}

%\markboth{}{}%
\maketitle
\thispagestyle{fancy}
\fancyhf{}
\renewcommand{\headrulewidth}{0pt}
\chead{\footnotesize This work has been submitted to the IEEE for possible publication. Copyright may be transferred without notice, after which this version may no longer be accessible.}
\begin{abstract}
Analysis of sub-synchronous oscillations (SSO) in IBR-dominated grids relies on frequency scan-based estimation of black-box IBR models at selected operating points. Since IBRs may operate over a wide range of operating conditions, frequency responses obtained at a limited number of operating points may not adequately represent the dynamics required for system-level SSO analysis. Accurate parameterization of operating-point-dependent IBR dynamics is challenging due to the heterogeneous dynamic behaviors that may arise across the operating space. This paper addresses this challenge by analytically characterizing the conditions that give rise to discontinuous and non-smooth variations in IBR dynamics. Leveraging these insights, a geometric representation based on singular value decomposition is used to identify coherent operating regions and partition the operating space into dynamically consistent regions. Within each region, the operating-point dependence of the IBR frequency response is accurately captured using simple linear regression. The proposed framework is validated on a modified IEEE 39-bus system. Results demonstrate that the parameterized IBR frequency responses accurately reconstruct system-level dynamics at the prevailing operating condition, enabling frequency-response and modal analysis without repeated system-level frequency scans.
\end{abstract}
\section{Introduction}
\IEEEPARstart{G}{rowing} share of inverter-based resources (IBRs) is reshaping system dynamics and introducing new grid stability challenges \cite{tf-definitions}. In particular, IBR-driven sub-synchronous oscillations (SSOs) have emerge as a difficult phenomenon to foresee, analyze and mitigate \cite{Gu2026, ESIG2024}. Several reported SSO events around the world highlight the growing operational challenges associated with IBR integration, as these oscillations often arise from adverse interactions between IBR controls and the network under specific operating conditions \cite{realworld, Modi2024, ESIG2024}. Consequently, there is increasing interest in methodologies capable of tracking operating-point-dependent system dynamics and providing early warning of SSO as operating conditions evolve.

Two main challenges towards systematic analysis of SSO are: 1) high-fidelity vendor-specific IBR models are opaque or black-box to system operators, and 2) IBR dynamics often exhibit operating-point dependence, causing potentially significant changes in their frequency-response characteristics across the operating space. A common approach to characterize black-box IBR dynamics is through dynamic frequency scans (DFS), which estimate the frequency response of the device at selected operating points \cite{ESIG2024,Cifuentes-Garcia-2026,Ramakrishna2023}. Where wide-area EMT models containing black-box , system-level DFS can be used to directly estimate the overall system frequency response at a given operating point \cite{NESO2025}. However, this approach is computationally intensive, requiring repeated wide-area EMT simulations across multiple frequencies for each operating condition of interest. Long settling times for poorly damped conditions and potential nonlinearities further increase the burden, rendering system-level DFS impractical for near real-time SSO analysis \cite{NESO2025}.

These limitations motivate a modular bottom-up approach in which system dynamics are reconstructed from individual IBR frequency responses. By combining operating-point-dependent IBR representations with the network (and other component models), the overall system dynamics can be updated as operating conditions change, without requiring repeated frequency scans on wide-area EMT models. This provides a scalable foundation for operating-point-dependent system assessment, enabling near real-time monitoring, early warning and mitigation strategies for IBR-driven oscillations \cite{Gu2026}.

A key requirement for such an approach is the ability to accurately represent the operating-point-dependency of IBRs dynamics. Since DFS can only be performed at a limited number of operating conditions across an IBR's operating range, suitable parametric representations are required to obtain the corresponding frequency responses throughout the operating space. This enables the reconstruction of operating-point-dependent system dynamics by combining individual IBR parametric representations with the network model, facilitating systematic frequency-response and modal analysis for IBR-driven SSO \cite{Cheah-Mane2026}.

Existing research on operating-point IBR parametric representations has primarily relied on interpolation techniques, regression models, and machine learning approaches to approximate the mapping between operating points and IBR frequency responses or transfer functions \cite{Li2024, data-driven-1, Nabil2024, Ge2026, Wu2025}. Representative examples include support vector machines and Gaussian process regression using operating variables such as terminal voltage, active power and reactive power \cite{Nabil2024, data-driven-1}; neural-network-based framework for black-box inverters with implicit or unknown dynamics \cite{Li2024}; deep neural networks for estimating sequence impedance models of aggregated IBR plants in the presence of measurement noise \cite{Ge2026}; and stacked autoencoder architectures for impedance estimation and extrapolation to unstable operating conditions \cite{Wu2025}. These methods have demonstrated reasonable accuracy and illustrate the growing interest in parameterization of IBR operating point dependence.

While these methods differ significantly in complexity, they share a common challenge: the underlying heterogeneity in IBR dynamics across its entire operating range. Variations in active and reactive power injections, terminal voltage, or controller settings modify the operating point and may alter the dominant dynamic characteristics. More generally, parameter variations in dynamical systems can induce structural changes in their modal characteristics \cite{Gallina2011}. As a result, different operating conditions may exhibit distinct frequency-response characteristics. Consequently, a single global parametric representation constructed across the entire operating range may attempt to fit data associated with fundamentally different dynamic behaviors, reducing estimation accuracy and interpretability while increasing model complexity \cite{Amsallem2016}.

A similar challenge is encountered in parameter-dependent linear time-invariant (LTI) systems, particularly in the context of model order reduction (MOR) frameworks \cite{Amsallem2012, Amsallem2016, Benner2015, Resch-Schopper2026,Ionita2014}. There, changes in parameter values may alter modal dominance, trigger modal interactions, or induce mode switching, resulting in distinct dynamic behaviors across the parameter space  \cite{Amsallem2016}. As a consequence, reduced-order models generated at different operating conditions may not represent the dominant dynamics, making the construction of a single global parametric representation difficult, and often inaccurate \cite{Benner2015, Amsallem2012, Amsallem2016}. A widely adopted approach is to partition the parameter space into regions exhibiting coherent dynamic behaviors, and construct local parametric representations within each region \cite{Amsallem2016, Benner2015}. By ensuring dynamic consistency within each subspace, relatively simple interpolation or regression techniques can often provide accurate representations of the underlying parameter-dependent dynamics \cite{Benner2015, Resch-Schopper2026}.

In this paper, we leverage this approach to identify coherent operating regions (CORs) directly from IBR frequency responses. First, an analytical characterization is derived to explain the conditions under which discontinuous or non-smooth variations emerge in IBR dynamics as operating point evolves. Building on this insights, a geometric representation of the singular value decomposition (SVD) of IBR frequency responses is introduced to capture the structural properties of the underlying input-output dynamics. The resulting geometric features enable the identification of operating regions exhibiting distinct dynamic behaviors. By constructing independent parametric representations within each region, the proposed framework avoids the inconsistencies associated with fitting a single global parametric representation.

The main contributions of this work are as follows:
\begin{itemize}
    \item[1)] An analytical formulation to characterize the conditions under which discontinuities or non-smooth operating-point dependence arises in IBR frequency responses or transfer functions.
    \item[2)] A geometric representation derived from the SVD of IBR frequency responses that captures both directional and gain-related properties of the associated input-output subspaces.
    \item[3)] A COR framework that partitions the operating space accordingly to underlying IBR dynamic behavior and enables accurate parameterization of IBR frequency responses using simple linear regression models.
\end{itemize}
\begin{figure}
    \centering
    \includegraphics[width=\linewidth]{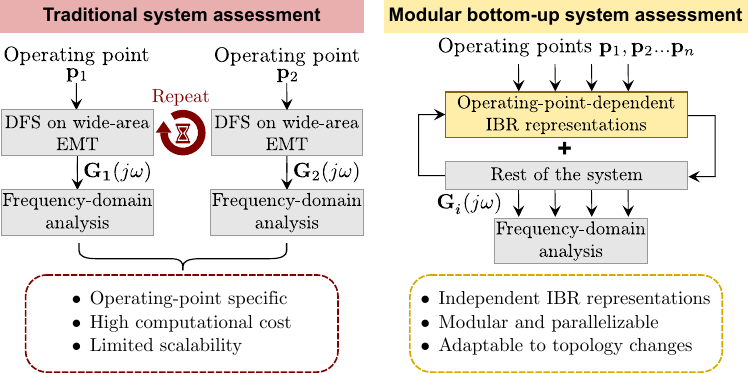}
    \caption{Traditional system-level assessment versus modular bottom-up reconstruction of operating-point-dependent dynamics.}
    \label{fig:main-diagram}
\end{figure}

Accurate parametric representation of IBRs operating-point-dependency enables the reconstruction of system dynamics by interconnecting individual devices frequency-response or transfer function representations with the network model. This eliminates the need for repeated wide-area DFS as operating conditions evolve, while naturally accommodating topology and operating-point changes through a modular framework. Consequently, the proposed approach provides a modular and scalable foundation for monitoring and early warning of IBR-driven SSO. Figure~\ref{fig:main-diagram} contrasts conventional wide-area DFS on EMT models with the proposed modular framework, illustrating how operating-point-dependent IBR representations enable bottom-up reconstruction of system dynamics for scalable system-level analysis.

\section{Geometric representation of operating-point-dependent dynamics} \label{sec:mathback} 
\subsection{Notations}
Let $\mathbb{R}$ and $\mathbb{C}$ denote the sets of real and complex numbers, respectively. The complex conjugate is denoted by $(\cdot)^*$. For a matrix $\mathbf{A}$, $\mathbf{A}^\mathsf{T}$ and $\mathbf{A}^H$ denote its transpose and conjugate transpose, respectively.  The Euclidean and Frobenius norms are denoted by $\|\cdot\|_2$ and $\|\cdot\|_F$, respectively, with the latter defined as $\|\mathbf{A}\|_F = \sqrt{\sum_{i,j} |a_{ij}|^2}$.  The identity matrix in $\mathbb{C}^{n\times n}$ is denoted by $\mathbb{I}_n$. For notational simplicity, the subscript $n$ is omitted when the dimension is clear from the context.
\subsection{Transfer function continuity and smoothness}
Consider an LTI parameter-dependent system, evaluated at $\mathbf{p}$ and described by the transfer function $\mathbf{G}(s, \mathbf{p})\!\in\!\mathbb{C}^{n\times n}$. Let $\mathbf{p}_0$ and $\mathbf{p}_1$ denote two operating points (setpoints) around which the system is linearized. To describe the variation of the system dynamics between these conditions, we introduce a convex combination of the operating space through a parameter $\alpha \in [0,1]$, defined as $\mathbf{p}_\alpha = (1-\alpha)\mathbf{p}_0 + \alpha \mathbf{p}_1$, under the assumption that the system remains stable along the entire trajectory. This parametrization defines a continuous trajectory between the two operating points, allowing intermediate system representations for continuity and smoothness analysis. For each value of $\alpha$, the transfer function can be written as
\begin{equation}
\mathbf{G}(s,\mathbf{p}_\alpha)=\mathbf{G}_\alpha(s) = \mathbf{C}_\alpha (s\mathbb{I} - \mathbf{A}_\alpha)^{-1} \mathbf{B}_\alpha ,
\end{equation}
where $\mathbf{D}_\alpha=0$. To evaluate how the dynamic response varies between two operating points, consider the difference between the corresponding transfer functions
\begin{align}
\Delta \mathbf{G}_\alpha(s)&=\mathbf{G}_\alpha(s)\!-\!\mathbf{G}_0(s)
= (\mathbf{C}_\alpha\!-\!\mathbf{C}_0)(s\mathbb{I}\!-\!\mathbf{A}_\alpha)^{-1}\mathbf{B}_\alpha \notag\\
& +\mathbf{C}_0\big[(s\mathbb{I}\!-\!\mathbf{A}_\alpha)^{-1}\!-\!(s\mathbb{I}-\mathbf{A}_0)^{-1}\big]\mathbf{B}_\alpha \notag\\
& + \mathbf{C}_0(s\mathbb{I}\!-\!\mathbf{A}_0)^{-1}(\mathbf{B}_\alpha\!-\!\mathbf{B}_0).
\end{align}
This decomposition separates the variations in the transfer response into contributions associated with the input/output matrices $\mathbf{B}$ and $\mathbf{C}$, and variations arising from the internal system dynamics $\mathbf{A}$. The latter can be further characterized using the matrix identity
$\mathbf{X}^{-1}-\mathbf{Y}^{-1}= \mathbf{X}^{-1}(\mathbf{Y}-\mathbf{X})\mathbf{Y}^{-1}$ 
for invertible matrices. Applying this relation yields
\begin{align}
\Delta \mathbf{G}_\alpha (s)
&= (\mathbf{C}_\alpha\!-\!\mathbf{C}_0)(s\mathbb{I}\!-\!\mathbf{A}_\alpha)^{-1}\mathbf{B}_\alpha \notag\\
&+\! \mathbf{C}_0 (s\mathbb{I}\!-\!\mathbf{A}_\alpha)^{-1}(\mathbf{A}_\alpha\!-\!\mathbf{A}_0)(s\mathbb{I}\!-\!\mathbf{A}_0)^{-1} \mathbf{B}_\alpha \notag\\
&+\! \mathbf{C}_0(s\mathbb{I}\!-\!\mathbf{A}_0)^{-1}(\mathbf{B}_\alpha\!-\!\mathbf{B}_0).
\end{align}
Under the assumption that both matrices $\mathbf{B}$ and $\mathbf{C}$ are non operating-point-dependent, the former expression can be reduced to \begin{equation}
    \Delta \mathbf{G}_\alpha(s)\approx \mathbf{C}_0 (s\mathbb{I}\!-\!\mathbf{A}_\alpha)^{-1}(\mathbf{A}_\alpha\!-\!\mathbf{A}_0)(s\mathbb{I}\!-\!\mathbf{A}_0)^{-1} \mathbf{B}_\alpha. \label{eq:BAC}
\end{equation}

Since $\mathbf{G}_\alpha(s)$ is stable for all $\alpha \in [0,1]$, the resolvent $\|(s\mathbb{I} - \mathbf{A}_\alpha)^{-1}\|_F$ is bounded. Consequently, the continuity of the transfer function with respect to $\alpha$ is governed by the variations of the state matrix $\mathbf{A}_\alpha$. To interpret these variations in terms of the underlying system dynamics, each state matrix can be expressed through its left and right eigenvector decomposition as $\mathbf{A}_i = \mathbf{\Psi}_i \mathbf{\Lambda}_i \mathbf{\Phi}_i^{\mathsf{T}}$. The difference between two operating conditions can then be written as
\begin{align}
    \mathbf{A}_\alpha\!-\!\mathbf{A}_0 &= \mathbf{\Psi}_\alpha\mathbf{\Lambda}_\alpha\mathbf{\Phi}_\alpha^{\mathsf{T}}-\mathbf{\Psi}_0\mathbf{\Lambda}_0\mathbf{\Phi}_0^{\mathsf{T}}\notag \\
    &= \mathbf{\Psi}_\alpha\left[\mathbf{\Lambda}_\alpha-\mathbf{\Phi}_\alpha^{\mathsf{T}}\mathbf{\Psi}_0 \mathbf{\Lambda}_0 \mathbf{\Phi}^{\mathsf{T}}_0\mathbf{\Psi}_\alpha\right]\mathbf{\Phi}_\alpha^{\mathsf{T}}\notag \\
    &= \mathbf{\Psi}_\alpha\left[\mathbf{\Lambda}_\alpha-(\mathbf{\Phi}^{\mathsf{T}}_0\mathbf{\Psi}_\alpha)^{-1} \mathbf{\Lambda}_0 (\mathbf{\Phi}^{\mathsf{T}}_0\mathbf{\Psi}_\alpha)\right]\mathbf{\Phi}_\alpha^{\mathsf{T}}.
\end{align}

This formulation reveals that the variation between the system matrices can be interpreted through the relative alignment between the modal directions ($\mathbf{\Phi}_0^{\mathsf{T}}\mathbf{\Psi}_\alpha$) associated at different operating points. Consequently, the continuity and smoothness of $\Delta \mathbf{G}_\alpha(s)$ with respect to the operating parameter $\alpha$ depend not only on the displacement of the eigenvalues, but also on changes in the orientation of the modal subspaces that define the system dynamics.

\begin{figure}[H]
\centering
\begin{minipage}[t]{0.32\columnwidth}
\definecolor{startCol}{RGB}{255,178,0} 
\begin{tikzpicture}[scale=0.8,>=Latex]
\draw[->, thick] (-2,0) -- (1,0) node[above]{Re};
\draw[->, thick] (0,-2) -- (0,2) node[above]{Im};
\foreach \i in {5,...,19}{
    \pgfmathsetmacro{\x}{-2+2*\i/20}
    \pgfmathsetmacro{\y}{0+sqrt(2*\i/20)}
    \fill[white, draw=startCol] (\x,\y) circle (1.5pt);
    \fill[white, draw=startCol] (\x,-\y) circle (1.5pt);
}
\draw[startCol!70!black, -stealth] 
    plot[samples=70, domain=0.1:1.1, variable=\t] 
    ({-2+2*\t}, {0+sqrt(2*\t)});
\draw[startCol!70!black, -stealth] 
    plot[samples=70, domain=0.1:1.1, variable=\t] 
    ({-2+2*\t}, {-0-sqrt(2*\t)});
\node at (0,-2.3){(a)};
\end{tikzpicture}
\end{minipage}\hfill
% Case 3
\begin{minipage}[t]{0.32\columnwidth}
\begin{tikzpicture}[scale=0.8,>=Latex]
\definecolor{startCol}{RGB}{72,170,72} 
\definecolor{startCol2}{RGB}{255,178,0} 

\draw[->, thick] (-2,0) -- (1,0) node[above]{Re};
\draw[->, thick] (0,-2) -- (0,2) node[above]{Im};
\foreach \i in {5,...,18}{
    \pgfmathsetmacro{\x}{-0.2 - \i/18}
    \pgfmathsetmacro{\y}{1.2*sqrt(1-(\i/18.1))}
    \draw[startCol] (\x,\y) circle (1.5pt);
    \draw[startCol2] (\x,-\y) circle (1.5pt);
}
% Dots on Re axis
\foreach \i in {0,1,2}{
    \pgfmathsetmacro{\x}{-1.65 + 0.3*\i}
    \fill[white, draw=startCol] (\x,0) circle (1.5pt);
}

\draw[startCol!70!black, -stealth, thick]
    (-1.2,0) -- (-1.85,0);

\draw[startCol2!70!black, -stealth, thick]
    (-1.2,0) -- (-0.35,0);

\draw[startCol!70!black, -stealth] 
    plot[samples=70, domain=0.1:0.99, variable=\t] 
    ({-0.2 - \t}, {1.2*sqrt(1-(\t)});
\draw[startCol2!70!black, -stealth] 
    plot[samples=70, domain=0.1:0.99, variable=\t] 
    ({-0.2 - \t}, {-1.2*sqrt(1-(\t)});

% Dots and trajectories on Re axis after coalescence
\foreach \i in {0,1}{
    \pgfmathsetmacro{\x}{-1.65 + 0.3*\i}
    \fill[white, draw=startCol] (\x,0) circle (1.5pt);
}
\foreach \i in {2,3}{
    \pgfmathsetmacro{\x}{-1.65 + 0.3*\i}
    \fill[white, draw=startCol2] (\x,0) circle (1.5pt);
}

\node at (0,-2.3){(b)};
\end{tikzpicture}
\end{minipage}\hfill
% Case 2
\begin{minipage}[t]{0.32\columnwidth}
\begin{tikzpicture}[scale=0.8,>=Latex]

\definecolor{startCol}{RGB}{0, 170, 170}
\definecolor{endCol}{RGB}{255,178,0} 

% Axes
\draw[->, thick] (-2.4,0) -- (1.4,0) node[above]{Re};
\draw[->, thick] (0,-2.2) -- (0,2.2) node[above]{Im};

\foreach \i in {-6,...,24}{
    \pgfmathsetmacro{\t}{\i/10} % now t in [-1.2,1.2]
    \pgfmathsetmacro{\x}{-2.1 + 1.2*(\t+1.2)/2.4}

    % Uncoupled lines
    \pgfmathsetmacro{\yA}{1.15 + 0.15*\t}   
    \pgfmathsetmacro{\yB}{0.55 + 0.55*\t}   
    \pgfmathsetmacro{\yc}{0.5*(\yA+\yB)}
    \pgfmathsetmacro{\yd}{0.5*(\yA-\yB)}
    \pgfmathsetmacro{\delta}{0.10}

    \pgfmathsetmacro{\y}{\yc + sqrt(\delta*\delta + \yd*\yd)}

    \draw[startCol] (\x,\y) circle (1.3pt);
    \draw[startCol] (\x,-\y) circle (1.3pt);
}
\draw[startCol!40!black]
    plot[samples=20, domain=-0.4:1.1, variable=\t]
    ({
      -2.1 + 1.2*(\t+1.2)/2.4
    },
    {
      (0.5*((1.15 + 0.15*\t) + (0.55 + 0.55*\t))
      + sqrt(0.10*0.10 + (0.5*((1.15 + 0.15*\t) - (0.55 + 0.55*\t)))^2))
    });
\draw[startCol!40!black, -stealth]
    plot[samples=20, domain=-0.8:2.7, variable=\t]
    ({
      -2.1 + 1.2*(\t+1.2)/2.4
    },
    {
      (0.5*((1.15 + 0.15*\t) + (0.55 + 0.55*\t))
      + sqrt(0.10*0.10 + (0.5*((1.15 + 0.15*\t) - (0.55 + 0.55*\t)))^2))
    });
\draw[startCol!40!black]
    plot[samples=20, domain=-0.8:2.7, variable=\t]
    ({
      -2.1 + 1.2*(\t+1.2)/2.4
    },
    {
      -(0.5*((1.15 + 0.15*\t) + (0.55 + 0.55*\t))
      + sqrt(0.10*0.10 + (0.5*((1.15 + 0.15*\t) - (0.55 + 0.55*\t)))^2))
    });
\draw[startCol!40!black, -stealth]
    plot[samples=20, domain=-0.8:2.7, variable=\t]
    ({
      -2.1 + 1.2*(\t+1.2)/2.4
    },
    {
      -(0.5*((1.15 + 0.15*\t) + (0.55 + 0.55*\t))
      + sqrt(0.10*0.10 + (0.5*((1.15 + 0.15*\t) - (0.55 + 0.55*\t)))^2))
    });
\foreach \i in {0,...,24}{
    \pgfmathsetmacro{\t}{\i/10}
    \pgfmathsetmacro{\x}{-2.1 + 1.2*(\t+1.2)/2.4}

    \pgfmathsetmacro{\yA}{1.15 + 0.15*\t}
    \pgfmathsetmacro{\yB}{0.55 + 0.55*\t}

    \pgfmathsetmacro{\yc}{0.5*(\yA+\yB)}
    \pgfmathsetmacro{\yd}{0.5*(\yA-\yB)}
    \pgfmathsetmacro{\delta}{0.10}

    \pgfmathsetmacro{\y}{\yc - sqrt(\delta*\delta + \yd*\yd)}

    \draw[endCol] (\x,\y) circle (1.3pt);
    \draw[endCol] (\x,-\y) circle (1.3pt);
}
\draw[endCol]
    plot[samples=20, domain=0:2.7, variable=\t]
    ({
      -2.1 + 1.2*(\t+1.2)/2.4
    },
    {
      (0.5*((1.15 + 0.15*\t) + (0.55 + 0.55*\t))
      - sqrt(0.10*0.10 + (0.5*((1.15 + 0.15*\t) - (0.55 + 0.55*\t)))^2))
    });
\draw[endCol!70!black, stealth-]
    plot[samples=20, domain=-0.4:2.7, variable=\t]
    ({
      -2.1 + 1.2*(\t+1.2)/2.4
    },
    {
      (0.5*((1.15 + 0.15*\t) + (0.55 + 0.55*\t))
      - sqrt(0.10*0.10 + (0.5*((1.15 + 0.15*\t) - (0.55 + 0.55*\t)))^2))
    });
\draw[endCol]
    plot[samples=20, domain=0:2.7, variable=\t]
    ({
      -2.1 + 1.2*(\t+1.2)/2.4
    },
    {
      -(0.5*((1.15 + 0.15*\t) + (0.55 + 0.55*\t))
      - sqrt(0.10*0.10 + (0.5*((1.15 + 0.15*\t) - (0.55 + 0.55*\t)))^2))
    });
\draw[endCol!70!black, stealth-]
    plot[samples=20, domain=-0.4:2.7, variable=\t]
    ({
      -2.1 + 1.2*(\t+1.2)/2.4
    },
    {
      -(0.5*((1.15 + 0.15*\t) + (0.55 + 0.55*\t))
      - sqrt(0.10*0.10 + (0.5*((1.15 + 0.15*\t) - (0.55 + 0.55*\t)))^2))
    });
\node at (0,-2.4){(c)};
\end{tikzpicture}
\end{minipage}
\caption{Eigenvalue trajectories under structural changes: (a) Hopf bifurcation, (b) Mode coalescence, (c) Mode veering.}
\label{fig:eigTraj3Cases}
\end{figure}
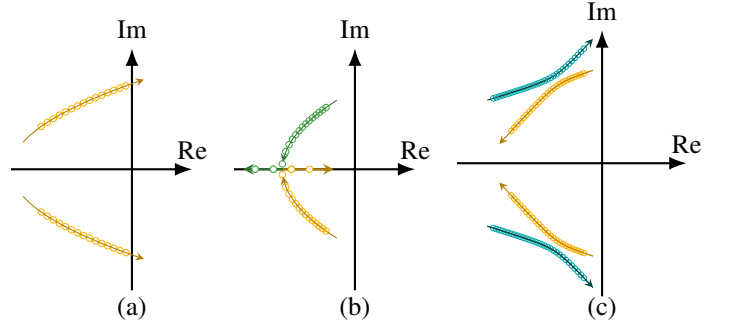

These effects are illustrated in Fig.~\ref{fig:eigTraj3Cases}, where different types of variations in eigenvalue trajectories and modal alignment are observed. In case (a), poles approaching the imaginary axis (Hopf bifurcation) significantly affect the smoothness of the resolvent $(s\mathbb{I}-\mathbf{A})^{-1}$. In cases (b) and (c), two modal interaction phenomena are illustrated, corresponding to mode coalescence and mode veering, respectively. The former is associated with the merging of modes, whereas the latter results from modes approaching each other and exchanging their dominant characteristics. In both cases, abrupt variations in the associated eigenvectors can be observed \cite{Gallina2011}.

\textbf{Remark 1.} Structural changes in the eigenvalues do not necessarily lead to discontinuities or non-smooth behaviors in $\mathbf{G}_\alpha(s)$. From (\ref{eq:BAC}), these properties are governed by both modal interactions and the input-output structure of the system. Therefore, an unobservable or uncontrollable mode cannot give rise to discontinuities or non-smooth behavior in the transfer function.

An analogous interpretation of system dynamic behavior can be obtained directly from frequency-response representations. Expressing the corresponding frequency-response matrices with the SVD as $\mathbf{G}_i=\mathbf{U}_i \mathbf{\Sigma}_i \mathbf{V}_i^H$, with $\!\mathbf{\Sigma}_i\!=\!\text{diag}(\sigma_1,\dots,\sigma_n)$, allows variations in the dynamic response across operating points to be decomposed in terms of their singular components as
\begin{align}
\Delta \mathbf{G}_\alpha(j\omega) &= \mathbf{U}_\alpha \mathbf{\Sigma}_\alpha \mathbf{V}_\alpha^{H}-\mathbf{U}_0 \mathbf{\Sigma}_0 \mathbf{V}_0^{H}\notag \\
&= \mathbf{U}_\alpha \Big[\mathbf{\Sigma}_\alpha-\underbrace{(\mathbf{U}_\alpha^H\mathbf{U}_0)}_{{U\text{-alignment}}}\mathbf{\Sigma}_0 \underbrace{(\mathbf{V}_\alpha^H \mathbf{V}_0)}_{V\text{-alignment}}\hspace{0pt}^H\Big] \mathbf{V}_\alpha^{H},
\end{align}
where the frequency dependence of $\mathbf{U}(j\omega)$, $\mathbf{\Sigma}(j\omega)$, and $\mathbf{V}(j\omega)$ is omitted for notation simplification. This expression highlights that variations in the frequency-response can be decomposed into three contributions: changes in the amplification of the system response $\mathbf{\Sigma}_\alpha\!-\!\mathbf{\Sigma}_0$ (change in energy), and misalignments in the left and right singular subspaces, quantified through the alignment matrices $\mathbf{U}_\alpha^H\mathbf{U}_0$ and $\mathbf{V}_\alpha^H\mathbf{V}_0$. When the singular subspaces are well aligned, these matrices approach the identity, and the variation is primarily driven by the local sensitivity of the singular values and the parameter variation (i.e., $\nabla_{\mathbf p}\mathbf{\Sigma}_\alpha\cdot \Delta\mathbf{p}$). Otherwise, off-diagonal contributions may arise, indicating increased mixing between singular modes and potential changes in the system's dynamic behavior.

Considering the above conditions, the continuity and smoothness of the transfer function cannot be guaranteed when changes in the operating condition lead to changes in the system's dynamic behavior. Consequently, when developing parametric representations, using data from a wide range of operating conditions may reduce its accuracy due to inconsistent dynamic behavior across incoherent operating regions. This suggests the need to identify boundaries in the operating space where such phenomena occur. To address this, the following subsection introduces a geometric representation of the frequency-response that captures the intrinsic dynamics associated with each operating point.
\subsection{Geometric SVD representation}
Analyzing the alignment of the singular subspaces across all operating conditions, i.e., evaluating $\mathbf{U}_\alpha^H \mathbf{U}_0$ (or $\mathbf{V}_\alpha^H \mathbf{V}_0$) for each linearization and frequency sample, leads to a high-dimensional problem of order $\mathcal{O}(n^2 f)$, where $n$ denotes the number of operating conditions and $f$ the number of frequency samples. This dimensionality makes the resulting representation difficult to interpret and limits its scalability. To overcome this issue, each operating condition is instead represented through trajectories defined by its SVD,  avoiding the need to compute all pairwise alignments explicitly.

For clarity, the following analysis focuses on the $2\times2$ case, i.e., $\mathbf{G}_\alpha(j\omega)\in\mathbb{C}^{2\times2}$. Nevertheless, the proposed construction can be generalized to transfer functions of arbitrary dimension through an analogous representation in a higher-dimensional space. The SVD decomposes the system response into an input subspace rotation $\mathbf{V}_\alpha(j\omega)$, a diagonal gain matrix $\mathbf{\Sigma}_\alpha(j\omega)$, and an output subspace rotation $\mathbf{U}_\alpha(j\omega)$. By decoupling the amplification, captured by $\mathbf{\Sigma}_\alpha(j\omega)$, from the directional information encoded in the input-output subspaces, the system behavior can be analyzed in terms of gain and direction independently.

\begin{figure}
    \centering
    \includegraphics[scale=0.85]{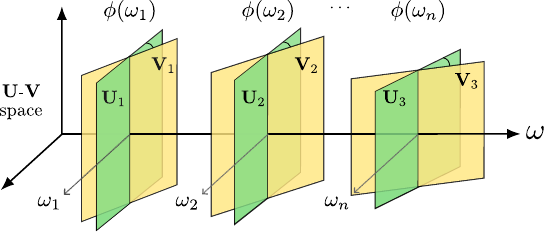}
    \caption{Geometric representation in the $\mathbf{U}\text{-}\mathbf{V}$ space. Each frequency $\omega_i$ defines an angle $\phi(\omega_i)$ between the subspaces spanned by the left (green) and right (yellow) singular vectors, enabling a representation of system alignment and identification of abrupt directional changes.}
    \label{fig:pangles}
\end{figure}

Let $\{\mathbf{u}_k(\omega)\}$ and $\{\mathbf{v}_k(\omega)\}$ denote the left and right singular vectors of $\mathbf{G}_\alpha(j\omega)$, respectively. The alignment between the input and output subspaces associated with the $k$-th singular value can be quantified through the angle between the subspaces spanned by the corresponding left and right singular vectors, defined as
\begin{equation}
    \cos\phi^{(k)}_{\alpha}(\omega) = |\mathbf{u}_k(\omega)^H\,\mathbf{v}_k(\omega)|.
    \label{eq:angle}
\end{equation}
Since $\{\mathbf{u}_k(\omega)\}$ and $\{\mathbf{v}_k(\omega)\}$ form orthonormal bases, both singular vector pairs define the same angle in the $2\times2$ case. Therefore, $\phi^{(1)}_{\alpha}\!=\!\phi^{(2)}_{\alpha}\!=\!\phi_{\alpha}$ represents a scalar measure of the relative orientation between input and output subspaces at each frequency. This is illustrated in Fig.~\ref{fig:pangles} for the $2\times2$ case, where each frequency $\omega_i$ defines a pair of subspaces spanned by $\mathbf{U}_i$ and $\mathbf{V}_i$, and their corresponding relative angle $\phi(\omega_i)$.

The angle $\phi_{\alpha}(\omega)$ provides a geometric interpretation as a measure of alignment between the dominant input and output directions of the system. When $\phi_{\alpha}(\omega)=0$, the corresponding singular vectors are aligned, and the direction of maximum input excitation is directly mapped to the direction of maximum output response, yielding a fully coherent amplification. In contrast, non-zero values of $\phi_{\alpha}(\omega)$ indicate a misalignment between the input and output subspaces, resulting in a loss of directional coherence in the input-output mapping.

In addition to the relative orientation captured by $\phi_{\alpha}(\omega)$, the distribution of energy across the singular modes provides complementary information on modal dominance. To quantify this, we define an anisotropy measure as\footnote{For the $2\times2$ case, $\eta_{\alpha}(\omega)$ can be interpreted as a normalized condition number, since $\eta_{\alpha}(\omega)=\left(\kappa(\omega)-1\right)/\left(\kappa(\omega)+1\right)$, where $\kappa(\omega)=\sigma_1(\omega)/\sigma_2(\omega)$.}
\begin{equation}
    \eta_{\alpha}(\omega) = \frac{\sigma_1(\omega) - \sigma_2(\omega)}{\sigma_1(\omega) + \sigma_2(\omega)}, \label{eq:eta}
\end{equation}
which quantifies the extent to which a single mode dominates the system response. Values of $\eta_{\alpha}(\omega)$ close to one indicate that the system response is strongly dominated by a single mode (or direction), whereas values close to zero correspond to a balanced contribution of multiple modes. Such conditions increase the potential for changes in the dominant system directions.

Thus, the dynamic behavior of each operating point $\mathbf{p}_\alpha$ can be characterized by the trajectories $\cos\phi_\alpha(\omega)$ and $\eta_\alpha(\omega)$. Similar trajectories across operating conditions indicate comparable input-output alignment and modal dominance, and consequently coherent dynamic behavior. In contrast, significant deviations reveal structural variations in the system dynamics, enabling the identification of coherent parametric regions. 

\textbf{Remark 2.} The quantities $\phi(\omega)$ and $\eta(\omega)$ depend only on the input-output behavior of the system, and are therefore invariant under similarity transformations of the state-space (i.e., $\!\mathbf{A}\!\rightarrow\!\mathbf{P}^{-1}\mathbf{A}\mathbf{P}$). This invariance reflects that the proposed descriptors are independent of the particular choice of state-space coordinates.

This geometric representation naturally extends to higher-dimensional systems by considering the alignment of multiple dominant singular modes through the angles between corresponding left and right singular vectors. Likewise, the anisotropy concept can be generalized through the hierarchical distribution of the singular values, enabling the characterization of modal dominance across higher-dimensional input-output subspaces.

These trajectories provide a comprehensive description of the system dynamics, capturing both the orientation and the modal dominance across the frequency range. This enables the analysis of operating-point-dependent behavior and the identification of coherent dynamical regimes. Consequently, resonances, transitions in modal dominance, and distinct dynamical behaviors can be identified through geometric patterns in the resulting trajectories, forming the basis of the proposed framework.
\section{Proposed Framework}\label{sec:methodology}
The interest in developing operating-point-dependent IBR representation lies in enabling the reconstruction of overall system dynamics under varying operating conditions. The reconstructed system model then provides a basis for frequency-response and modal analysis without requiring explicit access to the underlying device models, thereby enabling system-level analysis of IBR-driven SSOs. 

Building on the geometric representation introduced in Section~\ref{sec:mathback}, this section presents the proposed framework for operating-point-dependent representation of IBR dynamics, from power system modeling and COR identification to operating-point-dependent parameterization and system-level reconstruction.
\subsection{Power system modeling}
\begin{figure}
    \centering
    \includegraphics[width=\linewidth]{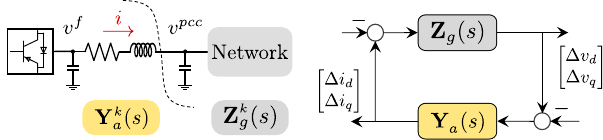}\\
    (a) \hspace{3.8cm} (b)
    \caption{Feedback system representation of the interaction between a device and the network: (a) physical system boundary, (b) mathematical feedback system.}
    \label{fig:boundaries}
\end{figure}

Power system components are represented in the $dq$ reference frame for small-signal analysis of IBR dynamics \cite{emt-rms}. The system operating point is characterized by $(p,q,v,\theta)$, which define the dependence of the linearized dynamics on the operating condition. IBRs are modeled through their admittance representations and connected to the network through a coupling impedance, as depicted in Fig.~\ref{fig:boundaries}(a). The remainder of the system is represented by an equivalent impedance $\mathbf{Z}_g(s)$, referred to as the Rest of the System (RoS), encompassing all dynamics not explicitly represented through individual device models. Here, $v_d$ and $v_q$ denote the PCC voltage components in the global $dq$ reference frame, while $i_d$ and $i_q$ are the corresponding injected current components. The input-output relationship for the $k$-th IBR is given by:
\begin{equation}
\begin{bmatrix}
\Delta i_{d,k}(s) \\
\Delta i_{q,k}(s)
\end{bmatrix}=
\underbrace{\begin{bmatrix}
Y_{dd}^k(s) & Y_{dq}^k(s) \\
Y_{qd}^k(s) & Y_{qq}^k(s)
\end{bmatrix}}_{=\mathbf{Y}_{a}^k(s)}
\begin{bmatrix}
\Delta v_{d,k}(s) \\
\Delta v_{q,k}(s)
\end{bmatrix},
\label{eq-dqframes}
\end{equation}
where $\mathbf{Y}_{a}^k(s)\!\in\! \mathbb{C}^{2\times2}$ denotes the admittance matrix of the $k$-th device. By defining the block-diagonal structure of all IBRs in the grid as $\mathbf{Y}_{a}(s)=\mathrm{diag}\left(\mathbf{Y}_{a}^1(s),\dots,\mathbf{Y}_{a}^n(s)\right)$, the overall system dynamics can be reconstructed through the feedback interconnection shown in Fig.~\ref{fig:boundaries}(b). This modular formulation enables the reconstruction of different overall system representations, $\mathbf{G}(s)$, depending on the selected voltage-current interconnection structure.

To obtain a correct overall system representation, all subsystem models must be expressed in a common global reference frame \cite{Gu2021, Cifuentes2022}, which ensures consistency of state variables and boundary conditions across devices. For the modeling of individual devices, it is often more convenient to derive their transfer functions in a local $dq$ reference frame. In this frame, the $d$-axis is aligned with the steady-state voltage at the PCC, i.e., $v_q = 0$, which is achieved by rotating the reference frame in the power-flow angle $\theta_{k}$.

Therefore, when reconstructing the overall system model from individually estimated IBR models, the corresponding rotation based on the operating point must be applied to map each local frame into the global reference frame \cite{Gu2021}:
\begin{align}
\mathbf{Y}_{a}^k(s) &= \mathcal{R}_{\theta_{k}}^{-1} \cdot \mathbf{Y}_{a,local}^k(s) \cdot \mathcal{R}_{\theta_k}, \label{eq:rot}\\
\mathcal{R}_{\theta_k} &=
\begin{bmatrix}
\cos\theta_k & \sin\theta_k \\
-\sin\theta_k & \cos\theta_k
\end{bmatrix}.
\end{align}
This formulation decouples the voltage phase dependence from the admittance model, rendering the local representation dependent only on $(p,q, v)$. In the remainder of this work, individual IBRs are assumed to be expressed in a local reference frame following this alignment convention.

\subsection{Identification of coherent operating regions (CORs)}
For each operating point $\mathbf{p}_i = \left[p_i, q_i, v_i\right]^{\mathsf{T}}$, the IBR admittance is characterized by the geometric representations (\ref{eq:angle}) and (\ref{eq:eta}). Although these quantities capture the relative orientation and modal dominance of the system, they do not explicitly account for variations in the overall response magnitude. In particular, the dominant gain, reflected by the maximum singular value, may vary significantly when an eigenvalue approaches the imaginary axis (illustrated in Fig.~\ref{fig:eigTraj3Cases}(a)). Accordingly, a compact feature vector is defined as
\begin{equation}
   \mathbf{f}(\mathbf{p}_i)= \left[\cos\phi_i(\omega), \eta_i(\omega), \text{log }\sigma_{1}(\omega)\right]^{\mathsf{T}}.
\end{equation}

Candidate CORs are obtained by applying a $K$-means algorithm to the resulting feature vectors. Here, $\log \sigma_1$ reduces the dynamic range of the dominant singular value, while $\cos\phi$ provides a smoother feature representation

The resulting clustering solutions are evaluated according to two complementary criteria: cluster separation and clustering robustness. First, robustness is quantified through the Adjusted Rand Index (ARI) \cite{Hubert1985}, computed from multiple resampled realizations of the dataset. This metric measures the consistency of the clustering and therefore provides an indication of the stability of the identified operating regions. Subsequently, cluster separation is assessed using the silhouette score \cite{Silhouettes}, which quantifies the degree of distinction between neighboring regions. The final COR partition is selected as the clustering solution achieving the highest silhouette score among those exhibiting a sufficiently robust classification.
\subsection{Parameterization of IBR operating-point dependency}
The IBR admittance representation consists of four complex components corresponding to the $dq$-frame entries. In this work, the real and imaginary parts of each component are estimated separately to avoid discontinuities caused by phase wrapping and magnitude changes associated with pole and zero displacements. To this end, eight multi-output linear regression models are fitted per IBR and COR, covering each frequency sample at which the admittance is estimated. For each IBR and COR, the regression models are fitted using the operating points $\{\mathbf{p}_k\}$ assigned to that region and their corresponding admittance responses $Y_{ij}^{(k)}(j\omega)$, with $i,j \in \{d,q\}$. For each $dq$-frame and frequency $\omega$, the regression problem is formulated as
\begin{align}
\left\{
\begin{aligned}
&\min_{\beta^{\text{Re}}} \sum_k\lVert \text{Re}(Y_{ij}^{(k)}(j\omega))\!-\!\beta_{ij}^{\text{Re}}(\omega)^{\mathsf{T}}\mathbf{p}_k\!-\! \beta_{0,ij}^{\text{Re}}(\omega)\rVert_2^2 \\
&\min_{\beta^{\text{Im}}} \sum_k\lVert \text{Im}(Y_{ij}^{(k)}(j\omega)) \!-\! \beta_{ij}^{\text{Im}}(\omega)^{\mathsf{T}}\mathbf{p}_k\!-\!\beta_{0,ij}^{\text{Im}}(\omega)\rVert_2^2
\end{aligned}
\right.
\end{align}
where $\!\beta_{ij}(\omega)\!\in\!\mathbb{R}^3\!$ and $\!\beta_{0,ij}(\omega)\!\in\mathbb{R}$ correspond to the regression coefficients at each frequency point. It is important to note that, although a linear regression model is used in this work, the methodology can be applied to more complex modeling approaches.
\begin{figure}
    \centering
    \includegraphics[width=\linewidth]{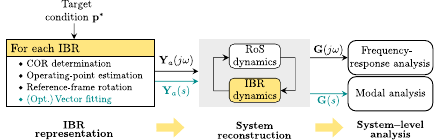}
    \caption{System reconstruction framework based on operating-point-dependent IBR parametric representations for system-level analysis.}
    \label{fig:methodology}
\end{figure}

The estimation of the admittance representation at an arbitrary operating point $\mathbf{p}_\alpha$ requires the identification of its corresponding COR. While this is straightforward for operating points located well inside a region, points near region boundaries may not be uniquely associated with a single cluster. To ensure a smooth transition across regions, the admittance can be expressed as a weighted combination of the local parametric representations associated with each COR. The weighting coefficients $w_i$ can be determined based on the proximity of $\mathbf{p}_\alpha$ to each region in the parameter space\footnote{A normalization step is required to ensure a consistent and comparable distance measure in the parameter space.}, as follows
\begin{align}
d_i(\mathbf{p}_\alpha)\!=\!\min_{\mathbf{p}_k\in\mathcal{P}_i}\|\mathbf{p}_\alpha\!-\!\mathbf{p}_k\|_2, \hspace{0.1cm}
    w_i\!=\! \frac{\exp(-d_i(\mathbf{p}_\alpha))}{\!\sum_{\ell\in\mathcal{C}}\exp(-d_\ell(\mathbf{p}_\alpha))\!}
\end{align}
where $\mathcal{C}$ denotes the set of CORs, and $\mathcal{P}_i$ the set of operating points of the $i$-th COR. This formulation enables a smooth transition between models and accounts for the ambiguity in region identification near the boundaries.
\vspace{-0.5em}\subsection{System-level reconstruction}
The complete reconstruction framework is illustrated in Fig.~\ref{fig:methodology}. The overall system frequency response can be reconstructed by interconnecting the operating-point-dependent IBR representations through the closed-loop formulation shown in Fig.~\ref{fig:boundaries}(b). Furthermore, vector fitting can be applied independently to each IBR representation, avoiding the need to identify a single high-order model of the overall system. The resulting state-space representations can then be interconnected through the same closed-loop formulation, enabling frequency-response and modal analysis. Since the operating-space partitioning is performed offline, online implementation only requires identifying the corresponding COR and evaluating its associated parametric representation.
\begin{table}
\centering
\caption{IBR configurations and setpoints in the modified IEEE 39-bus system. Power quantities are normalized on a 100~MVA base.}
\begin{adjustbox}{width=0.9\columnwidth,center}
\begin{tabular}{ccccc|cccc}
\hline
       & \multicolumn{4}{c}{Scenario I} & \multicolumn{4}{c}{Scenario II} \\ \hline
       Bus & Mode & $p$      & $q$      & $v$      & Mode & $p$       & $q$      & $v$      \\ \hline \hline
30  & \textcolor{black}{GFM$_1$}  & 3.38    & 0.58      & 0.99   & \textcolor{black}{GFM$_1$}   & 5.64       & -0.18      & 0.94      \\
31  & \textcolor{black}{GFM$_2$} & 9.77      & 1.31      & 0.94  & \textcolor{black}{GFM$_2$}    & 1.48       & 2.78     & 0.94     \\
32  & \textcolor{black}{GFM$_3$} & 6.50     & 1.49   & 0.95  & \textcolor{black}{GFM$_3$}    & 6.50     & 1.73      & 0.95     \\
33  & \textcolor{black}{GFM$_4$} & 6.25     & 0.50   & 0.97  & \textcolor{Green}{GFL$_1$}    & 9.37       & 0.38      & 0.96    \\
34  & \textcolor{black}{GFM$_5$} & 6.01      & 0.72     & 0.97  & \textcolor{Green}{GFL$_2$}    & 4.05       & 1.53      & 0.97      \\
35  & \textcolor{Green}{GFL$_1$} & 5.14      & -0.28      & 0.97  & \textcolor{Green}{GFL$_3$}    & 6.67       & 0.89      & 0.96      \\
36  & \textcolor{Green}{GFL$_2$} & 4.47      & 1.83      & 0.98  & \textcolor{Green}{GFL$_4$}    & 4.46      & 1.27     & 0.96  \\
37  & \textcolor{Green}{GFL$_3$} & 4.31      & 0.69      & 1.00 & \textcolor{black}{GFM$_4$}    & 7.30      & 0.26      & 0.97     \\
38  & \textcolor{Green}{GFL$_4$} & 6.61      & 0.07      & 1.03  & \textcolor{Green}{GFL$_5$}    & 8.98       & 0.46      & 1.01      \\
39  & \textcolor{Green}{GFL$_5$} & 8.82      & 2.18      & 0.95  & \textcolor{black}{GFM$_5$}    & 4.94      & 2.40   & 0.95 \\ \hline
\end{tabular}
\end{adjustbox}
\par\smallskip
\noindent(*) The subscript $i$ distinguishes different controller parameter sets considered in this work.
\label{tab:ops-ieee39}
\end{table}
\begin{figure}
    \centering
    \includegraphics[width=0.45\linewidth, trim={0.05cm 0.1cm 0.05cm 0.1cm},clip]{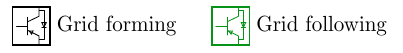} \\ \vspace{0.15cm}
    \includegraphics[width=0.49\linewidth]{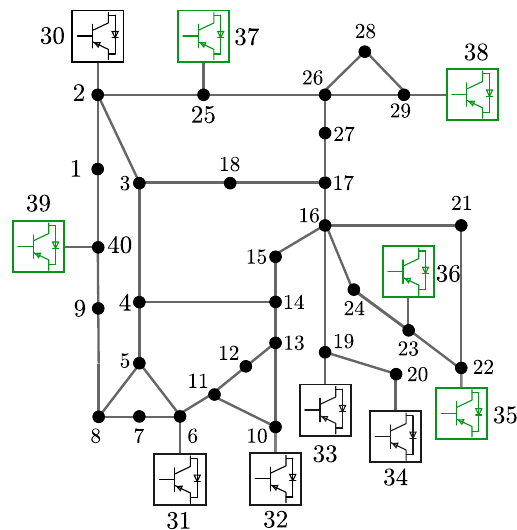}
    \includegraphics[width=0.49\linewidth]{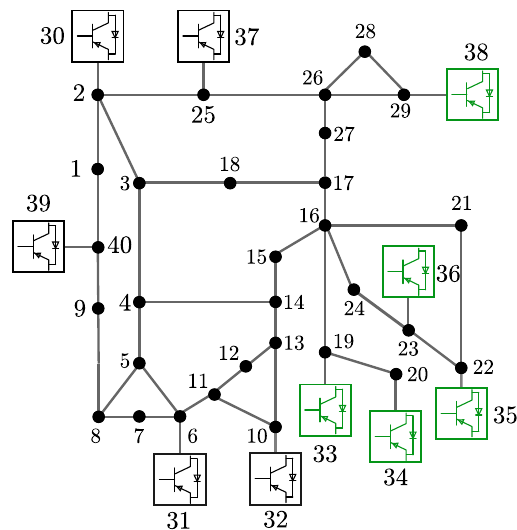}
    \caption{Modified IEEE-39 bus system with 10 IBRs: (left) scenario I, and (right) scenario II.}
    \label{fig:ieee-39}
\end{figure}
\vspace{-0.5em}\section{Case study}
The proposed framework is evaluated on a modified IEEE-39 bus system in which all synchronous machines are replaced by IBRs operating under different control modes. Minor modifications are also introduced to the original network configuration, including the addition of a coupling impedance at bus 39. The system is composed of five GFM and five GFL units with varying operating conditions, control parameters, and network locations, as depicted in Fig.~\ref{fig:ieee-39} and Table~\ref{tab:ops-ieee39}. Two scenarios are presented here to demonstrate the capability of the proposed methodology to estimate IBR admittance representations and reconstruct the overall system dynamics, supporting both frequency-response and modal analysis. 

For simplicity, and without loss of generality, the same uniform operating point sampling is adopted over $(p,q,v)$, with $p\in[0,1150]~\mathrm{MW}$, $q\in[-300,300]~\mathrm{MVar}$, and $v\in[0.93,1.07]~\mathrm{pu}$. The ranges are discretized into six equally spaced operating points for $p$, seven for $v$, and eight for $q$, resulting in 336 operating conditions per IBR.
\begin{figure*}
    \centering
    \includegraphics[width=\linewidth, trim={0 0cm 0 0},clip]{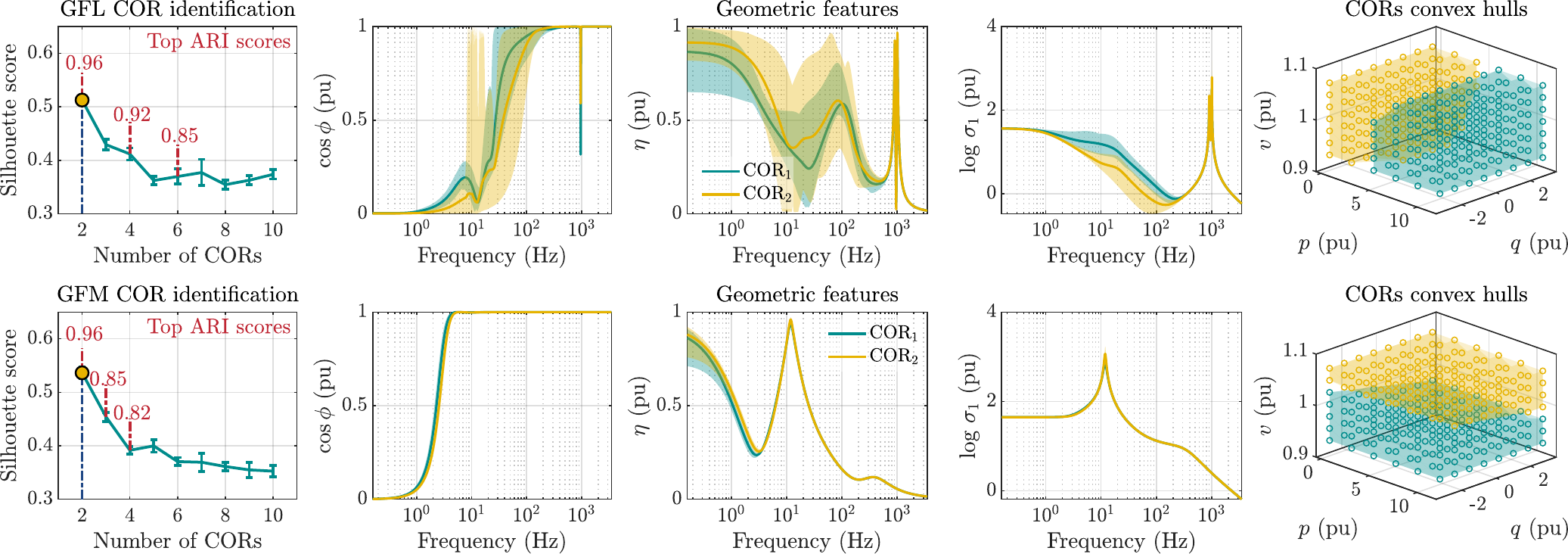} 
    \vspace{0mm}
    \makebox[\linewidth]{
        \makebox[0.2\linewidth][c]{\hspace{10mm}(a)}
        \makebox[0.2\linewidth][c]{}
        \makebox[0.2\linewidth][c]{(b)}
        \makebox[0.2\linewidth][c]{}
        \makebox[0.2\linewidth][c]{(c)}
    }
    \caption{Geometric feature-based COR identification for representative GFL$_2$ (top row) and GFM$_3$ (bottom row) units. (a) Silhouette scores with error bars and corresponding ARI values used for COR selection, (b) geometric feature trajectories ($\cos\phi,\eta,\log\sigma_1$), and (c) identified CORs in the $(p,q,v)$ space.}
    \label{fig:clusters}
\end{figure*}

All operating conditions are evaluated to obtain their corresponding admittance frequency responses. The geometric SVD representation is then used to partition the operating space into CORs and fit their corresponding local parametric representations. The proposed methodology is compared against a global representation fitted using the complete operating space. For model accuracy assessment, the error of an estimated admittance frequency response $\hat{\mathbf{Y}}_a^k(j\omega)$ is computed as
\begin{equation}
    e_a^k(\omega)= \dfrac{\| \mathbf{Y}_a^k(j\omega)-\hat{\mathbf{Y}}_a^k(j\omega)\|_F}{\| \mathbf{Y}_a^k(j\omega)\|_F}.
\end{equation}

To illustrate the proposed estimation framework, one representative GFM and one representative GFL unit are analyzed in detail, including the CORs identification and the estimation of their corresponding local representation at operating points within each COR. Both devices are represented through their admittance models, including synchronization loops, outer control loops, inner current loops, filter dynamics, and coupling impedances. The GFL unit operates under PV control, whereas the GFM unit adopts a synchronizing Pf-droop control scheme. Detailed representations of the control structures are presented in \cite{emt-rms}.

The estimated IBR admittances representations are subsequently interconnected through the network model for overall system reconstruction. The network lines are represented using their $\pi$-equivalent models, consisting of $RL$ series branches and shunt capacitances at both terminals, whereas the loads are modeled as constant impedances.

The two scenarios correspond to different loading conditions and operating setpoints, leading to distinct oscillatory behaviors. Scenario I exhibits a dominant oscillatory mode at 13.12 Hz with a damping ratio of 13\%, mainly associated with GFMs 2 and 3 located at buses 6 and 10, respectively. Scenario II presents two dominant oscillatory modes: the first at 13.11 Hz with a damping ratio of 12.7\%, again associated with GFMs 2 and 3, and the second at 5.77 Hz with a damping ratio of 3.4\%, mainly associated with the GFL units located at buses 19, 20, and 22.
\vspace{-1em}
\section{Results and discussion}\label{sec:results}
\subsection{Geometric CORs identification and performance validation}
The proposed COR identification framework is applied to all IBRs in the system. The resulting clustering statistics are summarized in Table~\ref{tab:sil-scores}, which reports the silhouette score, the corresponding ARI, and the number of identified CORs for each device. For all IBRs, the selected solution consists of two CORs, indicating that at least two distinct dynamic regimes can be identified within the considered operating space. Moreover, the consistently high ARI values demonstrate that the identified regions are robust with respect to dataset resampling. The silhouette scores indicate varying degrees of separation between the identified operating regions, with GFL units generally exhibiting more distinct clustering structures than GFMs, consistent with the more heterogeneous dynamic behavior observed for this control mode.

To illustrate the proposed geometric characterization and COR identification process, GFL$_2$ and GFM$_3$ are selected as representative examples. Figure~\ref{fig:clusters} presents (a) the clustering metrics used for COR selection, including the silhouette score and the ARI values associated with the best clustering candidates, (b) the envelopes of the geometric feature trajectories associated with each COR, and (c) the resulting partitioning of the operating space visualized through the convex hulls of the identified CORs. The upper row corresponds to GFL$_2$, whereas the lower row corresponds to GFM$_3$.

For the GFL unit, the extracted features reveal a highly heterogeneous dynamic behavior across the operating space. This variability is primarily captured through the evolution of $\phi(\omega)$ and $\eta(\omega)$, indicating significant changes in both directionality and modal dominance. In particular, COR$_2$ exhibits strong variations in the frequency range of 10-20~Hz, where abrupt changes in $\phi(\omega)$ are accompanied by pronounced deviations in $\eta(\omega)$, reflecting a stronger sensitivity to operating-point variations and potential mode interactions. In contrast, COR$_1$ displays a significantly smoother behavior across this frequency range, with more consistent alignment and modal characteristics.

These observations suggest that COR$_2$ is associated with operating conditions exhibiting stronger directional sensitivity, whereas COR$_1$ cluster corresponds to a more consistent and stable regime. This distinction is further supported by the singular value behavior, where distinct high- and low-energy regions emerge, corresponding to the two identified CORs. These geometric features ultimately translate into a clear partition of the operating space, primarily driven by the active power setpoint.
\begin{table}
\caption{Selected number of CORs and corresponding average clustering quality metrics for each IBR.}
\centering
\renewcommand{\arraystretch}{0.9}
\begin{adjustbox}{width=0.9\columnwidth}
\begin{tabular}{cccccc} \hline
     & GFL$_1$    & GFL$_2$    & GFL$_3$    & GFL$_4$    & GFL$_5$    \\ \hline \hline
Silhouette score   & 0.5167 & 0.5126 & 0.5135 & 0.5162 & 0.5135 \\
ARI & 0.9656    & 0.9605    & 0.9556   & 0.9681  & 0.9607  \\ 
CORs & 2    & 2    & 2   & 2  & 2  \\ \hline
     & GFM$_1$    & GFM$_2$    & GFM$_3$    & GFM$_4$    & GFM$_5$   \\ \hline \hline
Silhouette score   & 0.4369 & 0.4911 & 0.5364 & 0.5643 & 0.4979 \\
ARI & 0.9227    & 0.9748 & 0.9556   & 0.9580    & 0.9837  \\ 
CORs & 2    & 2    & 2   & 2  & 2  \\ \hline
\end{tabular}
\end{adjustbox}
\label{tab:sil-scores}
\end{table}

For the GFM unit, a different dynamic behavior is observed. All three geometric features exhibit smooth variations across the entire operating range, indicating a consistent directional structure with no abrupt changes. This contrasts with the heterogeneous behavior observed in the GFL case. Nevertheless, a distinction between operating conditions can still be identified through $\eta(\omega)$, where variations in the dominant singular value emerge in the low-frequency range, consistent with the operating-point dependency introduced by the droop control mechanism. As a result, the identified CORs are mainly influenced by the operating voltage level. However, as shown in Fig.~\ref{fig:clusters}(c), the separation boundary also exhibits a dependency on the active and reactive power setpoints, leading to a combined effect of $(p,q,v)$ on the resulting partition.

\begin{figure}[t]
    \centering
    \includegraphics[width=\linewidth]{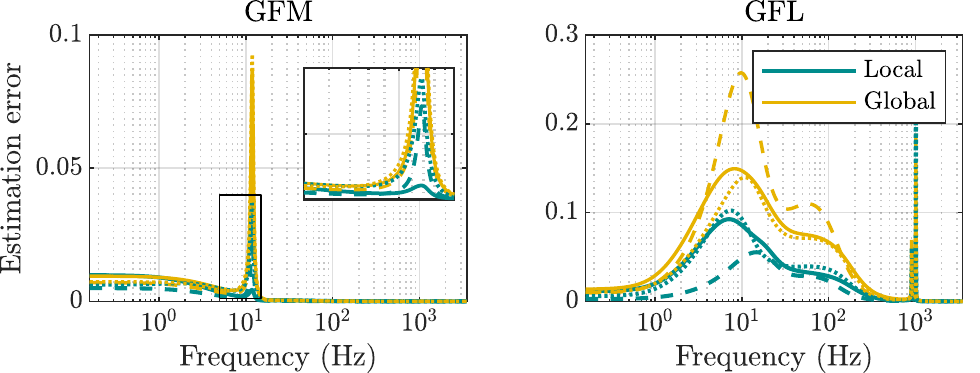} 
    \caption{Estimation error for GFM$_3$ (left) and GFL$_2$ (right) at $\mathbf{p}_1\!=\![6.2,-0.7,0.94]^{\mathsf{T}}$ (solid line),  $\mathbf{p}_2\!=\![1.5,1.2,1.02]^{\mathsf{T}}$ (dashed line), and $\mathbf{p}_3\!=\![4.5,2.7,0.99]^{\mathsf{T}}$ (dotted line)}
    \label{fig:estim_errors}
\end{figure}

To assess the effectiveness of the proposed operating-point-space partitioning on estimation accuracy, a global parametric representation (fitted using the complete dataset) is compared with a COR-based representation fitted using data from each COR. Three representative operating points are selected for validation: one from each identified COR ($\mathbf{p}_1$ and $\mathbf{p}_2$) and a third point ($\mathbf{p}_3$) located between both CORs. Figure~\ref{fig:estim_errors} shows the estimation error as a function of frequency for the GFM (left) and GFL (right) units, evaluated at $\mathbf{p}_1$, $\mathbf{p}_2$, and $\mathbf{p}_3$. 

For the GFM unit, the differences between the local and global representations are relatively small, reflecting the homogeneous geometric behavior observed previously. Nevertheless, the local representation consistently provides a lower estimation error around the dominant resonant frequency, as well as a slight improvement in the low-frequency range, where the operating-point dependency associated with the droop control mechanism is most pronounced. In contrast, the GFL unit exhibits a stronger dependence on the identified CORs. The local representation consistently outperform the global approximation for both CORs, with the most significant improvements observed at $\mathbf{p}_2$ (dashed lines). In both cases, the reduction in estimation error is concentrated around the frequency range where higher variability of $\eta(\omega)$ and $\phi(\omega)$ is observed in Fig.~\ref{fig:clusters}, although the magnitude of this reduction depends on the operating region. These results demonstrate that the proposed COR-based partitioning effectively captures operating regions with distinct dynamic behaviors and parameter sensitivities. As a result, local representation constructed within each COR provide a more accurate frequency-response estimation than a single global approximation.

Moreover, the results obtained at the boundary operating point $\mathbf{p}_3$ for both GFM$_3$ and GFL$_2$ remain consistent with the previous observations, with the local parametric representation systematically outperforming the global case. This suggests that the identified CORs successfully capture the dominant dynamic behaviors and parameter sensitivities within the operating space, even for operating conditions located close to the transition between neighboring regions.
\vspace{-1em}
\subsection{Modified IEEE-39 bus system reconstruction}
\begin{figure*}
    \centering
    \includegraphics[width=0.49\linewidth]{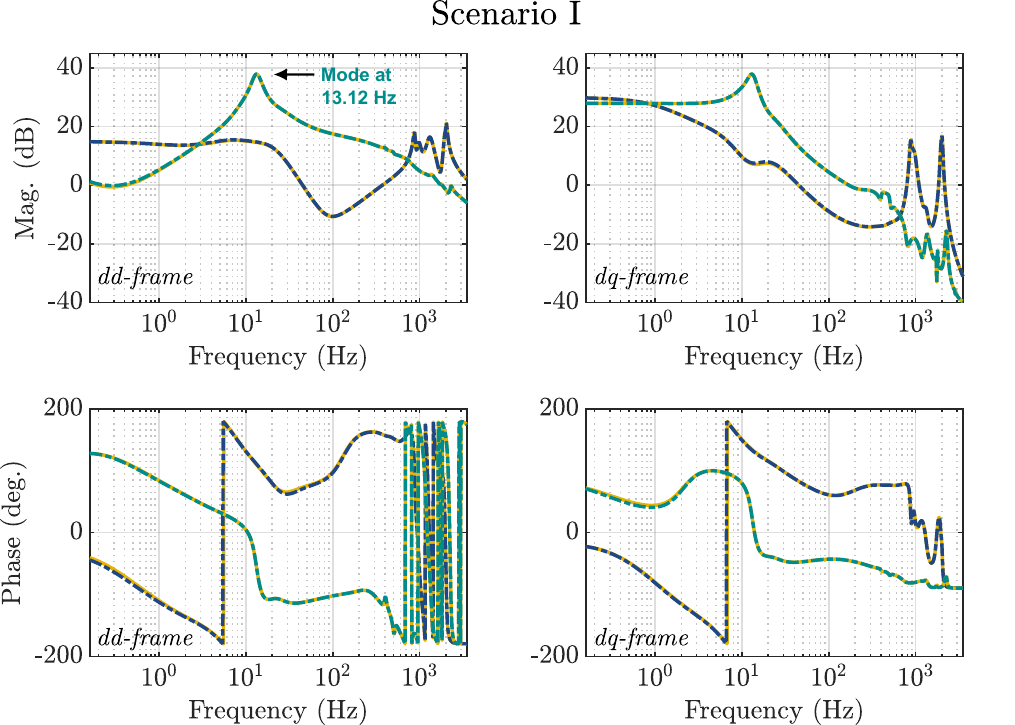}
    \includegraphics[width=0.49\linewidth]{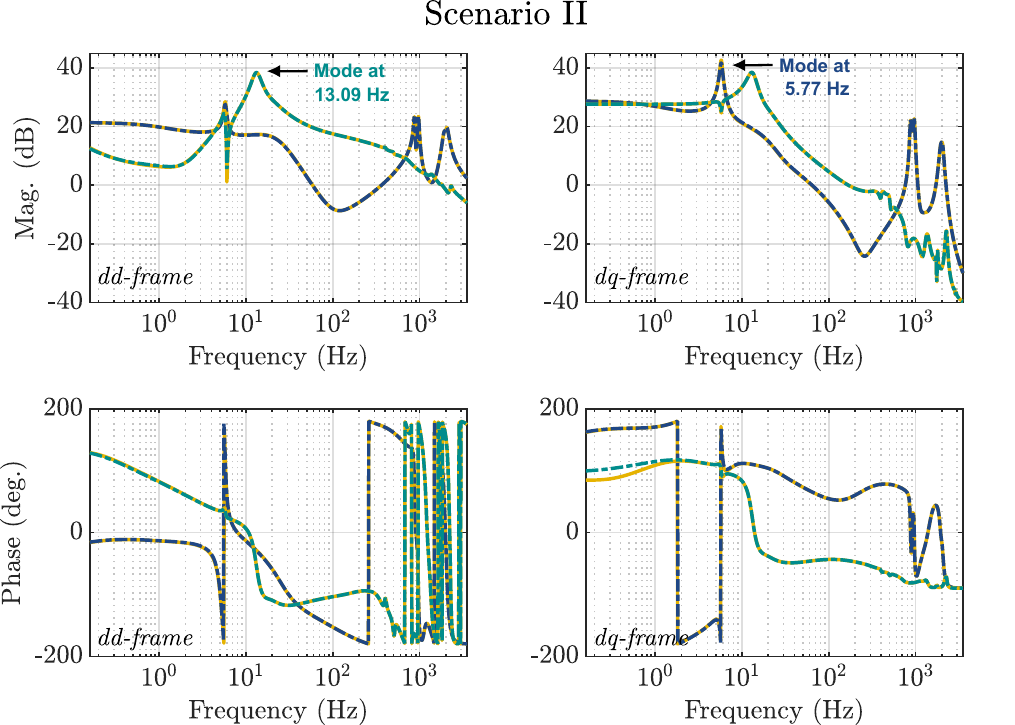}
    \caption{Overall system admittance reconstruction for the modified IEEE-39 bus system. Results at buses 6 (green) and 22 (blue) are compared with actual model-based values (yellow), for scenarios I (left) and II (right).}
    \label{fig:whole-sys}
\end{figure*}

The operating-point-dependent IBR representations obtained are then used to reconstruct the overall system dynamics for both scenarios. The resulting closed-loop admittance frequency responses are shown in Fig.~\ref{fig:whole-sys}. For clarity, only buses 6 and 22 are presented, corresponding to the locations of the IBRs with the highest participation factors in the dominant poorly damped oscillatory modes.

The reconstructed responses demonstrate high accuracy in capturing the dominant oscillatory modes in both scenarios. In Scenario I, the resonance around 13.12 Hz is accurately captured and is observable in both reference frames, with a dominant response at bus 6, where the corresponding GFM is connected. In Scenario II, the two dominant oscillatory modes are also identified. In particular, the mode at 13.09 Hz is captured in both frames, whereas the 5.77 Hz mode induced by the GFL exhibits higher gain in the $dq$-frame than in the $dd$-frame, reflecting both the device structure and the nature of the oscillatory interaction. A slight mismatch in the phase is nevertheless observed in the low-frequency range ($\leq1$ Hz) for bus 22. Despite this, the estimated overall system reconstruction remains highly accurate in capturing IBR-driven oscillatory dynamics.

\begin{figure}
    \centering \includegraphics[width=0.49\linewidth]{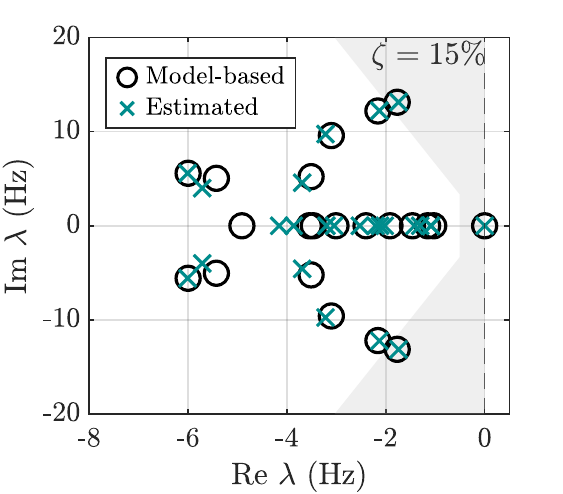}
    \includegraphics[width=0.49\linewidth]{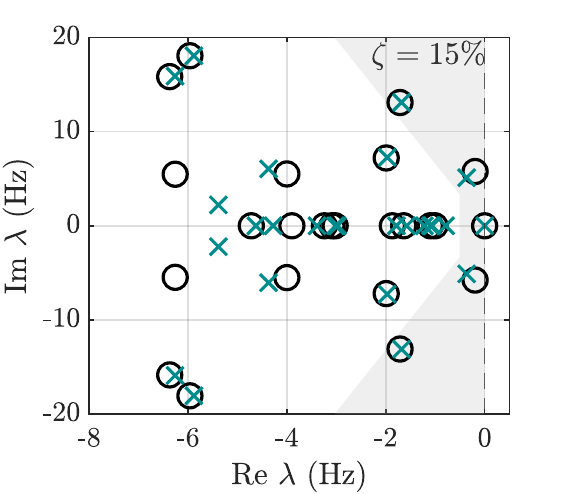}
    \caption{Estimated and model-based eigenvalues for Scenarios I (left) and II (right).}
    \label{fig:eigenvalues-models}
\end{figure}

To further extend the applicability of the proposed framework, vector fitting is applied to each estimated IBR frequency-response to obtain an equivalent transfer function. The resulting state-space models are subsequently interconnected to reconstruct the closed-loop for modal analysis. The eigenvalues obtained from the estimated overall system model are presented in Fig.~\ref{fig:eigenvalues-models}. Focusing on the frequency range of interest, it can be observed that the proposed framework accurately captures the poorly damped oscillatory modes in both scenarios, supporting its applicability for overall system analysis.

\begin{figure*}
    \centering
    \includegraphics[width=0.24\linewidth]{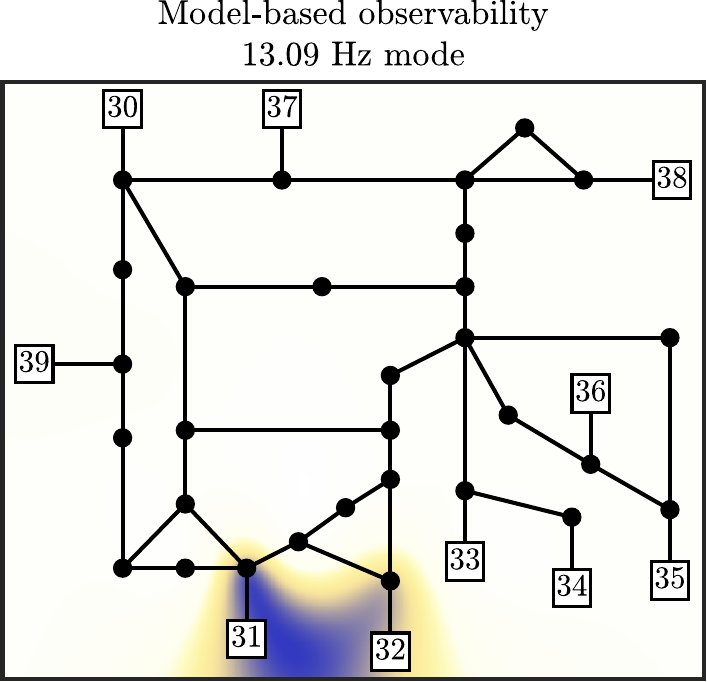}
    \includegraphics[width=0.24\linewidth]{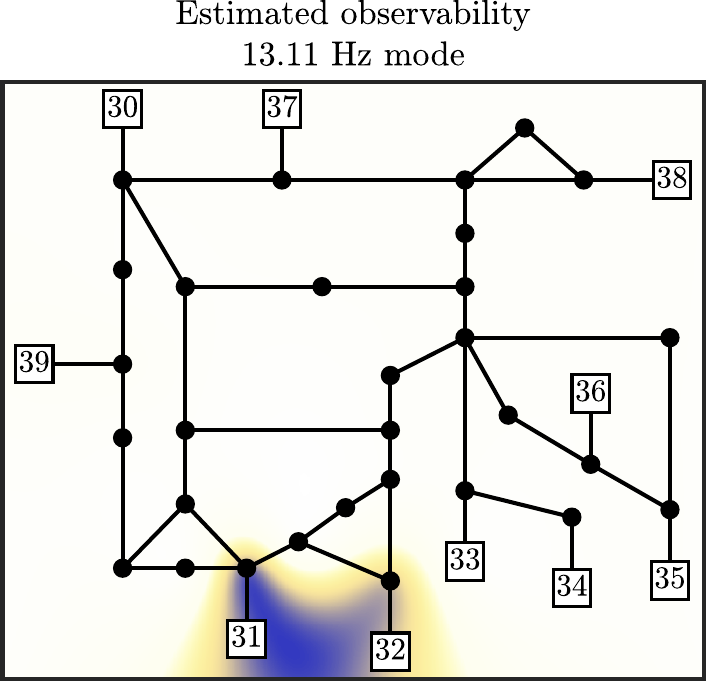} \hspace{0.1cm}
    \includegraphics[width=0.24\linewidth]{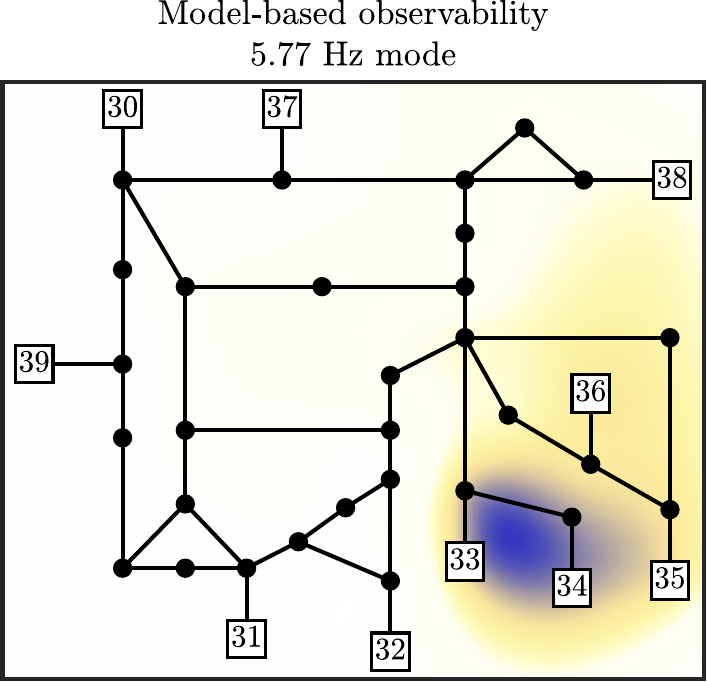}
    \includegraphics[width=0.24\linewidth]{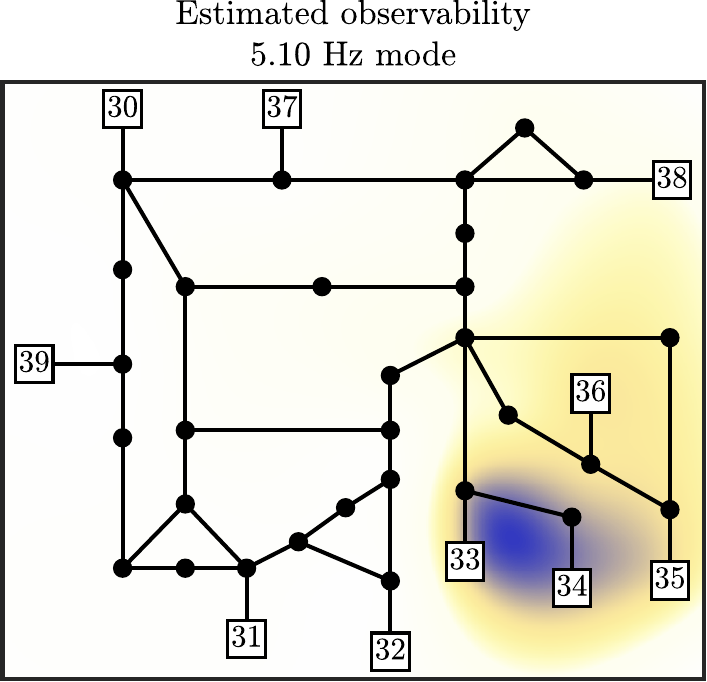} \\ \vspace{0.2cm}
    \includegraphics[scale=0.4]{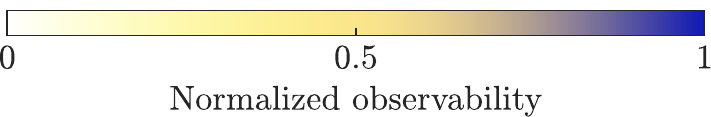}
    \caption{Estimated and model-based observability heatmaps for the two poorly damped modes in Scenario II, obtained from the overall system impedance.}
    \label{fig:heatmaps-applications}
\end{figure*}

Beyond modal analysis, Fig.~\ref{fig:heatmaps-applications} presents the normalized observability heatmaps of the two poorly damped modes identified in Scenario II, obtained using both the estimated and model-based system representations. In both cases, the estimated models accurately reproduce the dominant observability patterns, demonstrating that the reconstructed system dynamics preserve the key modal characteristics of the original system. This analysis can be readily extended to controllability, residues, and other modal metrics.
\vspace{-1em}
\subsection{Discussion}
This work has shown that, when using CORs, simple local parametric representations can accurately estimate IBR admittance representations across the operating space. However, the estimation accuracy depends not only on the selected modeling technique, but also on the characteristics of the system itself and on how representative the operating dataset is. Therefore, correctly characterizing each COR becomes essential. The results presented in this work show that different control modes may exhibit significantly different dynamic behaviors across the operating space. While some configurations behave relatively uniformly, others may present stronger operating-point-dependent dynamics, requiring a more detailed representation of the operating space. Consequently, rather than focusing only on reducing the number of operating points, greater attention should be placed on selecting operating conditions that properly capture the relevant system dynamics.
To achieve this, operating-point sampling strategies (such as \cite{Resch-Schopper2026}) capable of capturing the relevant system dynamics can enable the development of more accurate and less biased models. This highlights the importance of using sampling criteria that adapt the selection of operating conditions according to the dynamic variability of the system. Such approaches could also reduce uncertainty associated with estimation errors, enabling formal guarantees regarding the validity and accuracy of the reconstructed models across the considered operating ranges, while improving the identification of transition regions where changes in the dynamic structure occurs. Ultimately, these criteria could be incorporated into compliance and validation procedures, defining the operating points at which black-box IBR models should be evaluated before being delivered to system operators, allowing these models to be systematically exploited within frameworks such as the one presented in this work.

An important aspect highlighted by this work is that, once coherent operating regions are identified, relatively simple modeling techniques can accurately capture the operating-point dependency of IBR dynamics. In this context, the use of simple local representations not only improves interpretability, but also preserves smoothness and continuity properties that may be difficult to guarantee with more complex black-box estimators. More importantly, the identified CORs provide regions in which the sensitivity of the frequency response to operating-point variations exhibits a consistent and predictable behavior. This enables a clearer understanding of how changes in active power, reactive power, or voltage affect the underlying dynamics, facilitating applications such as sensitivity analysis, optimization, and corrective control design.
\section{Conclusions}
This paper demonstrates that the operating-point dependency of IBR dynamics can be accurately parametrized for system-level analysis of IBR-driven SSOs. The key enabler is identifying CORs across the entire operating range of an IBR. Within each COR, simple regression models accurately capture variations in IBR frequency-response or transfer function while preserving smoothness, interpretability, and computational efficiency. The results show that accurate parameterization depends not only on the operating conditions themselves, but also on the ability to identify and group operating points according to the heterogeneity in underlying dynamic behavior.

Building on this insight, the proposed framework is shown to enable a modular, bottom-up reconstruction of operating point-dependent system dynamics from black-box IBR mod- els. By combining parameterization of IBR operating point dependence with the rest of the grid model, the overall system frequency response and modal characteristics can be accurately obtained without repeating computationally intensive system- level dynamic frequency scans on a wide-area EMT model as operating conditions evolve. This significantly reduces computational burden while retaining the fidelity and accuracy required for early warning of incipient SSO and future mitigation strategies.
\vspace{-1.3em}
\bibliography{bibtex/bib/IEEEexample}
%\appendix

\end{document}